\documentclass[12pt]{article}

\pdfoutput=1
\catcode`\@=11
\@addtoreset{equation}{section}

\global\arraycolsep=2pt 
\usepackage{mathrsfs, amsbsy, amssymb, latexsym, amsfonts, amscd, amsmath, cite} 
\usepackage{graphicx,color}

\newcommand{\no}{\nonumber}

\allowdisplaybreaks

\begin{document}

\begin{flushright}
% \parbox{4cm}
\begin{tabular}{l}
{DIAS-STP-15-03} \\
{KUNS-2564}
\end{tabular}
\end{flushright}

\vspace*{1cm}

\begin{center}
{\Large\bf Chaotic strings in a near Penrose limit of AdS$_5\times T^{1,1}$}
\vspace*{2cm}\\ 
{\large 
Yuhma Asano$^{\ast}$\footnote{E-mail:~yuhma@stp.dias.ie}, 
Daisuke Kawai$^{\dagger}$\footnote{E-mail:~daisuke@gauge.scphys.kyoto-u.ac.jp},
Hideki Kyono$^{\dagger}$\footnote{E-mail:~h$\_$kyono@gauge.scphys.kyoto-u.ac.jp},
and Kentaroh Yoshida$^{\dagger}$\footnote{E-mail:~kyoshida@gauge.scphys.kyoto-u.ac.jp}} 
\end{center}
\vspace*{0.25cm}
\begin{center}
$^{\ast}${\it School of Theoretical Physics, Dublin Institute for Advanced Studies, \\
10 Burlington Road, Dublin 4, Ireland.}
\vspace*{0.2cm} \\ 
$^{\dagger}${\it Department of Physics, Kyoto University \\ 
Kyoto 606-8502, Japan.} 
\vspace*{0.25cm}
\end{center}
\vspace{0.5cm}

\begin{abstract}
We study chaotic motions of a classical string in a near Penrose limit of AdS$_5\times T^{1,1}$\,. 
It is known that chaotic solutions appear on $R\times T^{1,1}$\,, depending on initial conditions. 
It may be interesting to ask whether the chaos persists even in Penrose limits or not. 
In this paper, we show that sub-leading corrections in a Penrose limit provide an unstable separatrix, 
so that chaotic motions are generated as a consequence of collapsed Kolmogorov-Arnold-Moser (KAM) tori. 
Our analysis is based on deriving a reduced system composed of two degrees of freedom 
by supposing a winding string ansatz. 
Then, we provide support for the existence of chaos by computing Poincar\'e sections. 
In comparison to the AdS$_5\times T^{1,1}$ case, we argue that no chaos lives 
in a near Penrose limit of AdS$_5\times$S$^5$\,, 
as expected from the classical integrability of the parent system. 
\end{abstract}

\setcounter{footnote}{0}
\setcounter{page}{0}
\thispagestyle{empty}

\newpage

\tableofcontents

\section{Introduction}

AdS/CFT dualities \cite{M,GKP,W} are one of the most important subjects in string theory. 
The study of them diverges and continues to influence various fields including cosmology, 
nuclear physics, condensed matter physics, and recently non-linear dynamics. 
The most well-studied example is the duality between type IIB string theory on AdS$_5\times$S$^5$ 
and the $\mathcal{N}=4$ $SU(N)$ super Yang-Mills (SYM) theory in four dimensions. A recent progress 
is the discovery of an integrable structure behind this duality \cite{review}. The integrability has played 
an important role in checking the duality in non-BPS regions.  

\medskip 

In connection with the integrable structure, type IIB string theory on AdS$_5\times$S$^5$ is 
classically integrable in the sense that the Lax pair exists \cite{BPR}. 
Apart from this integrable example, there are many non-integrable AdS/CFT dualities 
in which chaotic string solutions appear. 
For example, when the internal space is given by a Sasaki-Einstein manifold 
like $T^{1,1}$ \cite{T11} and $Y^{p,q}$ \cite{Ypq}, the string world-sheet theory exhibits the chaotic behavior 
(For other examples, see \cite{AdS-soliton,D-brane,complex-beta,NR}). 

\medskip 

On the other hand, apart from the fundamental strings, chaotic motions of D0-branes 
in the BFSS matrix model \cite{BFSS} 
and the BMN matrix model \cite{BMN} have been shown in \cite{chaos-BFSS} 
and \cite{chaos-BMN}, respectively\footnote{For earlier works on chaos in classical (deformed) 
Yang-Mills theories, see \cite{YM,deformed-YM}.}. Thanks to the
mass-deformation, the BMN matrix model was robustly discussed by
computing Poincar\'e sections, which explicitly exhibit chaos.
It would be nice to consider a gravitational (or string theoretical) 
interpretation of the chaotic behavior of D0-branes. 
It may be related to a non-linear dynamical generation of space-time, 
fast scrambling of black hole \cite{SS} and the inequality proposed in \cite{MSS}. 
In fact, a fast thermalization in the BMN matrix model is discussed in \cite{Berenstein}. 

\medskip 

In this paper, we are concerned with non-integrable AdS/CFT dualities.  
As a particular example, we will concentrate on the AdS$_5\times T^{1,1}$ case, where  
the existence of chaos has been confirmed both numerically \cite{T11} and analytically \cite{Ypq}. 
The chaos appears basically because the classical string action on $R \times T^{1,1}$ contains 
a double pendulum as a subsystem. 
It may be interesting to ask whether the chaos persists even in Penrose limits \cite{Penrose} or not. 
The leading part in the limits gives rise to a free massive world-sheet theory. 
Then the sub-leading correction can be regarded as a small perturbation, but it is not so simple 
because quartic-order terms of canonical momenta are contained. 
Nevertheless, it is still possible to employ the standard procedure. 
Our analysis is based on deriving a reduced system composed of two degrees of freedom 
by supposing a winding string ansatz. 
Then, we provide support for the existence of chaos by computing 
Poincar\'e sections. In comparison to the AdS$_5\times T^{1,1}$ case, 
we argue that no chaos lives in a near Penrose limit of AdS$_5\times$S$^5$\,,  
as expected from the classical integrability of the parent system. 
 
\medskip 

The organization of this paper is as follows. 
In section 2, we consider a Penrose limit of AdS$_5\times T^{1,1}$ including the sub-leading corrections. 
In section 3, the bosonic light-cone Hamiltonian is derived on the near pp-wave background. 
The sub-leading corrections induce interaction terms in the system.  
In section 4, we show that chaotic string solutions exist in the resulting Hamiltonian system 
by computing Poincar\'e sections. 
In section 5, we revisit a near Penrose limit of AdS$_5\times$S$^5$ 
and argue that no chaos appears. 
Section 6 is devoted to conclusion and discussion.

\section{A near Penrose limit of AdS$_5\times T^{1,1}$} 

In this section we will consider a Penrose limit of the AdS$_5\times T^{1,1}$ background, 
including the sub-leading corrections. First of all, the metric of AdS$_5\times T^{1,1}$ is introduced in 
Sec.\ 2.1. Then we consider a Penrose limit of this background in Sec.\ 2.2. 

\subsection{The metric of AdS$_5\times T^{1,1}$}

Let us introduce the metric of the AdS$_5\times T^{1,1}$ background. 
The internal compact space $T^{1,1}$ is a five-dimensional Sasaki-Einstein manifold. 
The $T^{1,1}$ geometry is obtained as a base space of conifold 
(which is a Calabi-Yau three-fold) \cite{Candelas}.  
The AdS$_5\times T^{1,1}$ background is obtained as the near-horizon limit 
of a stack of $N$ D3-branes sitting at the tip of the conifold and the resulting geometry is considered 
as the gravity dual for an $\mathcal{N}=1$ superconformal field theory in four dimensions \cite{KW}. 

\medskip 

The metric of {AdS$_5\times T^{1,1}$} is given by 
\begin{eqnarray}
ds^2&=& R^2 (ds^2_{\rm AdS_5} + ds^2_{T^{1,1}})\,, \\ 
ds^2_{\rm AdS_5} &=&-\cosh^2 \rho\, dt^2 + d\rho^2 + \sinh^2 \rho\, d\Omega_3^2\,,  \\
ds^2_{T^{1,1}} &= & \frac{1}{9}\left[d\psi + \cos\theta_1\, d\phi_1+ \cos\theta_2\, d\phi_2\right]^2
+ \frac{1}{6}\sum_{i=1}^2\left[d\theta_i^2 + \sin^2\theta_i\, d\phi_i^2\right]\,.
\end{eqnarray}
Here $R$ is the radius of AdS$_5$\,. 
The isometry is $SU(2)_{\rm A} \times SU(2)_{\rm B} 
\times U(1)_{\rm R}$\,. 
Note here that $T^{1,1}$ is a homogeneous space and can be represented by the following coset\footnote{
In some references, the coset is said to be 
\[
\frac{SU(2)_{\rm A}\times SU(2)_{\rm B}}{U(1)}\,.
\]
However, this coset does not lead to the correct metric, as argued in the original paper \cite{Candelas}. 
The coset in (\ref{coset}) can reproduce the metric correctly and 
even three-parameter deformations of $T^{1,1}$ \cite{CO} as shown in \cite{CMY}.}: 
\begin{eqnarray}
T^{1,1} = \frac{SU(2)_{\rm A} \times SU(2)_{\rm B} 
\times U(1)_{\rm R}}{U(1)_{\rm A} \times U(1)_{\rm B}}\,. 
\label{coset}
\end{eqnarray}
Although the full Green-Schwarz string action has not been constructed yet, 
one may employ the bosonic part. In the following, we will concentrate on the bosonic part 
and consider classical string solutions moving on $R \times T^{1,1}$\,.

\subsection{A Penrose limit of AdS$_5\times T^{1,1}$} 

It is known that classical strings moving on $R \times T^{1,1}$ exhibit random motions i.e., chaos. 
Now we would like to consider a question, ``Can one observe chaos even in a near Penrose limit?''   
The answer is yes, as we will show later. 
Let us here introduce a near pp-wave geometry of AdS$_5\times T^{1,1}$ by including the sub-leading corrections 
in taking a Penrose limit. The leading part of the pp-wave geometry was originally discussed in \cite{IKM,GO,ZS}. 

\medskip 

To take a Penrose limit, a null-geodesic has to be picked up at first. 
Among the geodesics, we focus upon the {$\psi + \phi_1 + \phi_2$} direction in $T^{1,1}$\,.  
Then the light-cone coordinates 
$\tilde{x}^{\pm}$ and new angle variables $\Phi_i~(i=1,2)$ are introduced as\footnote{
Here the light-cone convention is slightly different from the one in \cite{IKM,GO,ZS}. 
Our convention follows the work \cite{Schwarz,Swanson} in which the sub-leading corrections 
are discussed in a near Penrose limit of AdS$_5\times$S$^5$\,. The present choice in (\ref{lc}) 
is convenient to deal with the sub-leading part. 
} 
\begin{eqnarray}
\tilde{x}^+ & \equiv & t\,, \quad \tilde{x}^- \equiv -t+\frac{1}{3}\left(\psi + \phi_1 + \phi_2\right)\,, 
\quad 
\Phi_1  \equiv \phi_1-t\,, \quad \Phi_2 \equiv \phi_2-t\,. \label{lc}
\end{eqnarray}
Then let us rescale the above coordinates by $R$ as follows:
\begin{equation}
\tilde{x}^+=x^+\,, \quad \tilde{x}^-=\frac{x^-}{R^2}\,, \quad \rho=\frac{r}{R}\,, 
\quad \theta_i = \sqrt{6}\, \frac{r_i}{R}\,.
\end{equation}
Finally, the $R \to \infty$ limit is taken. This is the Penrose limit we consider.  

\medskip 

After all, the resulting metric is given by 
\begin{eqnarray}
\label{expanded metric}
ds^2&=&{ds_0}^2+\frac{1}{R^2}\, {ds_2}^2+ {\cal O}\left( \frac{1}{R^4}\right)\,, \no \\
{ds_0}^2&=&2{dx^+}{dx^-}-\left({r}^2 + {r_1 }^2+{r_2}^2 \right) \left({dx}^+ \right)^2 
+ {dr}^2 +  {r}^2 {d\Omega_3}^2 \no \\ 
&& \qquad +{dr_1 }^2 + {r_1}^2 {d\Phi_1}^2
+{dr_2 }^2  +{r_2}^2 {d\Phi_2}^2\,,  \no \\
{ds_2}^2&=&\left(-\frac{1}{3}r^4+2{r_1 }^2 {r_2 }^2\right)\left(dx^+\right)^2 
-2\left({r_1 }^2+{r_2 }^2 \right){dx^+}{dx^-}
+\left(dx^-\right)^2+\frac{1}{3}{r }^4{d\Omega_3}^2 \no \\
&& \qquad +{r_1 }^2\left( 
-{r_1 }^2+2{r_2 }^2 \right) {dx^+}{d\Phi_1}
+{r_2 }^2\left( -{r_2 }^2+2{r_1 }^2 \right) {dx^+}{d\Phi_2}
-2 {r_1 }^2 {dx^- }{d\Phi_1}\no \\
&& \qquad -2 {r_2 }^2 {dx^- }{d\Phi_2}
+2{r_1 }^2{r_2 }^2{d\Phi_1}{d\Phi_2}
-{r_1 }^4{d\Phi_1 }^2
-{r_2 }^4{d\Phi_2 }^2
\,. \label{metric}
\end{eqnarray}
The leading part $ds_0^2$ is nothing but the familiar maximally supersymmetric 
pp-wave background \cite{BFHP}. 
As a matter of course, the sub-leading part $ds_2^2$ is different from that of AdS$_5\times$S$^5$\,. 
The sub-leading part $ds_2^2$ plays an important role in our later argument and 
indeed leads to chaotic string motions.

\section{Hamiltonian of a near pp-wave string} 

In this section, we will derive the light-cone Hamiltonian of a string moving 
on the near pp-wave background (\ref{metric})\,. 
Our derivation follows the procedure developed in \cite{Schwarz, Swanson} 
for the AdS$_5\times$S$^5$ case, 
though we employ only the bosonic part. 

\medskip 

We first work on a general background and solve the constraint conditions. 
Then the metric (\ref{metric}) is substituted into the resulting expression and 
the light-cone Hamiltonian we consider is derived.

\subsection{A light-cone string on a general background} 

Let us consider a general background with the metric 
$g_{\mu\nu}~(\mu,\nu = +,-,1\ldots,8)$ that satisfies the following conditions 
\[ 
g_{+I}=g_{-I}=0 \quad (I=1,\ldots,8)\,.
\] 
In addition, we suppose that 
the dilaton is constant and the NS-NS two-form is zero. 

\medskip 

The bosonic part of the classical string action is given by  
\begin{eqnarray}
\label{action}
{\cal S}_{\rm B}=\int\!d\tau d\sigma\,{\cal L}
=\frac{1}{2}\int\! d\tau d\sigma\,h^{ab}{\partial_a}x^{\mu}{\partial_b}x^{\nu}g_{\mu\nu}\,.
\end{eqnarray}
The string world-sheet is parametrized by $\tau$ and $\sigma$ and 
the dynamical variables $x^{\mu}(\tau,\sigma)$ describe the string dynamics. 
The quantity $h^{ab}~(a,b=\tau,\sigma)$ is defined as  
\[
h^{ab} \equiv \sqrt{-\gamma}\, \gamma^{ab}\,, \qquad \gamma \equiv \det(\gamma_{ab})\,, 
\] 
where $\gamma_{ab}$ is the world-sheet metric.

\medskip 

Then the canonical momenta $p_{\mu}$ are introduced as usual: 
\begin{equation}
\label{canonical momenta}
p_{\mu} = \frac{\partial \cal L}{\partial\left({\partial}_{\tau}x^{\mu}\right)}
=h^{\tau a}\partial_{a}x^{\nu}g_{\mu\nu}\,.
\end{equation}
Solving (\ref{canonical momenta}) in terms of ${\dot{x}}^{\mu}$ leads to the relation: 
\begin{equation}
\label{dotx}
{\dot{x}}^{\mu}=\frac{1}{h^{\tau\tau}}g^{\mu\nu}p_{\nu}-\frac{h^{\tau\sigma}}{h^{\tau\tau}}{x'}^{\mu}\,.
\end{equation}
Here the following notations have been introduced:
\[
{\dot{x}}^{\mu} \equiv \partial_{\tau}x^{\mu}\,, \qquad {x'}^{\mu} \equiv \partial_{\sigma}x^{\mu}\,. 
\]

\medskip 

The equation of motion for $h_{ab}$ provides constraint conditions,  
by which the energy-momentum tensor $T^{ab}$ is forced to vanish:
\begin{eqnarray}
T^{ab}=h^{ac}h^{bd}{\partial_c{x^{\mu}}}{\partial_d{x^{\nu}}}g_{\mu\nu}
-\frac{1}{2}h^{ab}h^{cd}{\partial_c{x^{\mu}}}{\partial_d{x^{\nu}}}g_{\mu\nu}=0\,. 
\label{Tmn}
\end{eqnarray}
By making use of (\ref{dotx})\,, $\dot{x}$ can be removed from the expression (\ref{Tmn}).  
Then the constraints in (\ref{Tmn}) can be written in terms of $p_{\mu}$ and ${x'}^{\mu}$ like   
\begin{eqnarray}
\label{constraint1}
p_{\mu}p_{\nu}g^{{\mu}{\nu}}+{x'}^{\mu}{x'}^{\nu}g_{\mu\nu}&=&0\,, \\
\label{constraint2}
p_{\mu}{x'}^{\mu}&=&0\,.
\end{eqnarray}

\medskip 

Let us here impose the light-cone gauge,  
\[
x^+= \tau\,, \qquad p_-={\rm constant}\,.
\]
Then the light-cone Hamiltonian $\mathcal{H}_{\rm lc}$ is defined as 
\[
\mathcal{H}_{\rm lc} \equiv - p_+\,.
\]
With (\ref{constraint1}) and (\ref{constraint2})\,, 
${x'}^-$ and $\mathcal{H}_{\rm lc}$ can be expressed in terms of $x^{I}$ and $p_{I}$\,:
\begin{eqnarray}
{x'}^-&=&-\frac{{x'}^{I}{p_{ I}}}{p_-}\,, \\  \label{light cone Hamiltonian}
{\cal{H}}_{\rm lc} &=&-\frac{p_-g^{+-}}{g^{++}}-\frac{1}{g^{++}} \sqrt{{p_-}^2g-g^{++}
\left(g_{--}\left(\frac{p_{{I}}{x'}^{{I}}}{{p_-}^2}\right)^2+p_{{I}}p_{{J}}g^{IJ}+{x'}^{{I}}{x'}^{{J}}g_{{I}{J}}\right)}\,,
\end{eqnarray}
where the following quantity has been introduced:
\[
g={(g^{+-})}^2-g^{++}g^{--}\,. 
\] 
Note that in the above derivation we have assumed that $g_{--}\neq 0$\,. 
When $g_{--}=0$ like the usual pp-wave metric (i.e., only the leading part), 
the light-cone Hamiltonian becomes 
\begin{eqnarray}
\label{light cone Hamiltonian g_--=0}
{\cal{H}}_{\rm lc}=-\frac{p_-g^{--}}{2g^{+-}}-\frac{1}{2g^{+-}p_-}\left(p_{I}p_{J}g^{IJ}+{x'}^{I}{x'}^{J}g_{IJ}\right)\,.
\end{eqnarray}
For a given metric, the expressions of ${\cal{H}}_{\rm lc}$ in (\ref{light cone Hamiltonian}) 
and (\ref{light cone Hamiltonian g_--=0}) are very useful.

\subsection{The Hamiltonian in the near pp-wave limit of AdS$_5\times T^{1,1}$}

For later argument, let us explicitly write down the light-cone Hamiltonian 
on the near pp-wave background (\ref{expanded metric}). 

\medskip 

By substituting the pp-wave metric (\ref{expanded metric}) into the formula (\ref{light cone Hamiltonian})\,, 
the light-cone Hamiltonian ${\cal{H}}_{\rm lc}$ is given by  
\begin{eqnarray}
{\cal{H}}_{\rm lc}&=&{\cal{H}}_0+\frac{1}{R^2}{\cal{H}}_{\rm int} + \mathcal{O}\left(\frac{1}{R^4}\right)\,, \label{LCH} \\
{\cal{H}}_0&=&\frac{1}{2}\left({p_r}^2+{p_{r_1}}^2+{p_{r_2}}^2+\frac{{p_{\Phi_1}}^2}{{r_1}^2}+\frac{{p_{\Phi_2}}^2}{{r_2}^2}
\right.\no\\
&& \left.+r^2+{r_1}^2+{r_2}^2+{r'}^2+{{r_1}'}^2+{{r_2}'}^2+{r_1}^2{{\Phi_1}'}^2+{r_2}^2{{\Phi_2}'}^2\right)\,, 
\label{T1,1 Hamiltonian 0} \\
{\cal{H}}_{\rm int}&=&-\frac{1}{8}\left({p_r}^2+{p_{r_1}}^2+{p_{r_2}}^2+\frac{{p_{\Phi_1}}^2}{{r_1}^2}
+\frac{{p_{\Phi_2}}^2}{{r_2}^2}-r^2+{r_1}^2+{r_2}^2+{r'}^2+{{r_1}'}^2+{{r_2}'}^2\right.\no \\
&& +\left.{r_1}^2{{\Phi_1}'}^2+{r_2}^2{{\Phi_2}'}^2\right)^2 
+\frac{1}{2}\left({p_{\Phi_1}}-{p_{\Phi_2}}\right)^2
-\frac{1}{2}\left({r_1}^2{{\Phi_1}'}-{r_2}^2{{\Phi_2}'}\right)^2 \no \\ 
&& +\frac{1}{6}r^4
+\frac{1}{2}\left({r_1}^4+{r_2}^4\right) 
+\frac{1}{2}\left({p_r}r'+{p_{r_1}}{r_1}'+{p_{r_2}}{r_2}'+{p_{\Phi_1}}{\Phi_1}'+{p_{\Phi_2}}{\Phi_2}'\right)^2\,. 
\label{T1,1 Hamiltonian int}
\end{eqnarray}
Here we have set $p_-=1$ and a constant term has been dropped off. 
In addition, we have ignored the terms concerned with $d\Omega_3^2$ for simplicity. 
In our later argument, we are not interested in the  angular part of AdS$_5$\,. 
In fact, the terms with $d\Omega_3^2$ can be dropped off by supposing that 
a constant position is taken on the S$^3$\,. 
In the following, we will not consider the higher-order terms with $\mathcal{O}(1/R^4)$ as well. 
Note also that $p_{\Phi_1}$ and $p_{\Phi_2}$ are constants of motion. 

\medskip 

The resulting system (\ref{LCH}) can be regarded as a sum of simple harmonic oscillators in ${\cal{H}}_0$ 
and a small perturbation by ${\cal{H}}_{\rm int}$\,. 
Hence it seems likely that the system is simple, but this is not the case actually. 
The interaction Hamiltonian ${\cal{H}}_{\rm int}$ contains four-order terms of canonical momenta 
and hence the Hamiltonian dynamics is quite intricate. Thus the behavior of classical trajectories 
is far from obvious and it is worth to study it. 

\medskip 

In the next section, we will consider the Hamiltonian dynamics with (\ref{LCH})
and show that chaotic string solutions are contained.

\section{Chaos in a near Penrose limit of AdS$_5\times T^{1,1}$} 

In this section, we show chaotic string motions in the near pp-wave background (\ref{expanded metric}). 
There are some standard methods to display chaotic motions (For an introductory book, see \cite{Strogatz}). 
Here we compute Poincar\'e sections\footnote{One may think of that Lyapunov exponents may be computed as well. 
However, it seems quite difficult to compute them in the present case as we will explain later.}.  
As evidence of chaos, the resulting sections show random motions with some islands. 

\medskip 

We study classical trajectories of a string moving on the near pp-wave background (\ref{expanded metric})\,. 
The light-cone Hamiltonian (\ref{LCH}) is very intricate and hence 
it is helpful to impose an ansatz to make the system much simpler. 
In addition, the string world-sheet is two-dimensional. 
Hence it is convenient to perform a dimensional reduction to one dimension 
by supposing a string wrapping on Cartan directions. 

\medskip 

Concretely, we consider two cases of a winding string. 
The one is that all of the motions are confined into the $T^{1,1}$ 
geometry. The other is that the radial direction of AdS$_5$ is included in the motions. 
In the following, we will investigate each of them. 

\subsection{Chaos in the $T^{1,1}$ directions}

The first ansatz we consider is the following:
\begin{eqnarray}
  \label{eq:ansatz_t11}
  & r = 0\,, \quad  p_r = 0\,,  \quad r_1 = r_1(\tau)\,, \quad \ p_{r_1} = p_{r_1}(\tau)\,, \quad 
r_2 = r_2(\tau)\,, \quad \ p_{r_2} = p_{r_2}(\tau)\,,  \no \\
  & \Phi_1 = \alpha_1\, \sigma\,, \quad p_{\Phi_1} = 0\,, \quad \Phi_2 = \alpha_2\, \sigma\,, 
\quad p_{\Phi_2} = 0\,.
\end{eqnarray}
Here $\alpha_i~(i=1,2)$ is an integer due to the periodicity of $\Phi_i$\,. 
This ansatz describes a string moving only in the $T^{1,1}$ geometry.  
Note that the spatial direction of the string world-sheet is wrapped on Cartan directions. 

\medskip 
 
Then the ansatz (\ref{eq:ansatz_t11}) reduces 
the free part (\ref{T1,1 Hamiltonian 0}) and 
the interaction part (\ref{T1,1 Hamiltonian int}) 
into the following forms: 
\begin{eqnarray}
  \mathcal{H}_0 &=& \frac{1}{2}\Bigl[ p_{r_1}^2 + p_{r_2}^2 
+ (1 + \alpha_1^2)\, r_1^2 + (1 + \alpha_2^2)\, r_2^2 \Bigr]\,, \nonumber \\ 
 \mathcal{H}_{\rm int} &=& - \frac{1}{8} \Bigl[p_{r_1}^2 + p_{r_2}^2 + (1 + \alpha_1^2)\,r_1^2 
+ (1 + \alpha_2^2)\,r_2^2 \Bigr]^2 \no \\
  && - \frac{1}{2} \left(\alpha_1\, r_1^2 - \alpha_2\, r_2^2 \right)^2 
+ \frac{1}{2} \left(r_1^4 + r_2^4\right)\,,
\label{hamiltonian_with_ansatz}
\end{eqnarray}
respectively. Now the dynamical variables of this system depend only on $\tau$\,, 
and it is simple enough to compute Poincar\'e sections. 

\medskip

In the following, 
we will provide numerical results to support the existence of chaos 
even in the near Penrose limit.

\subsubsection*{Poincar\'e section}

\begin{figure}[htbp]
\vspace*{0.1cm}
\begin{center}
\begin{tabular}{cc}
\includegraphics[scale=.25,angle=-0]{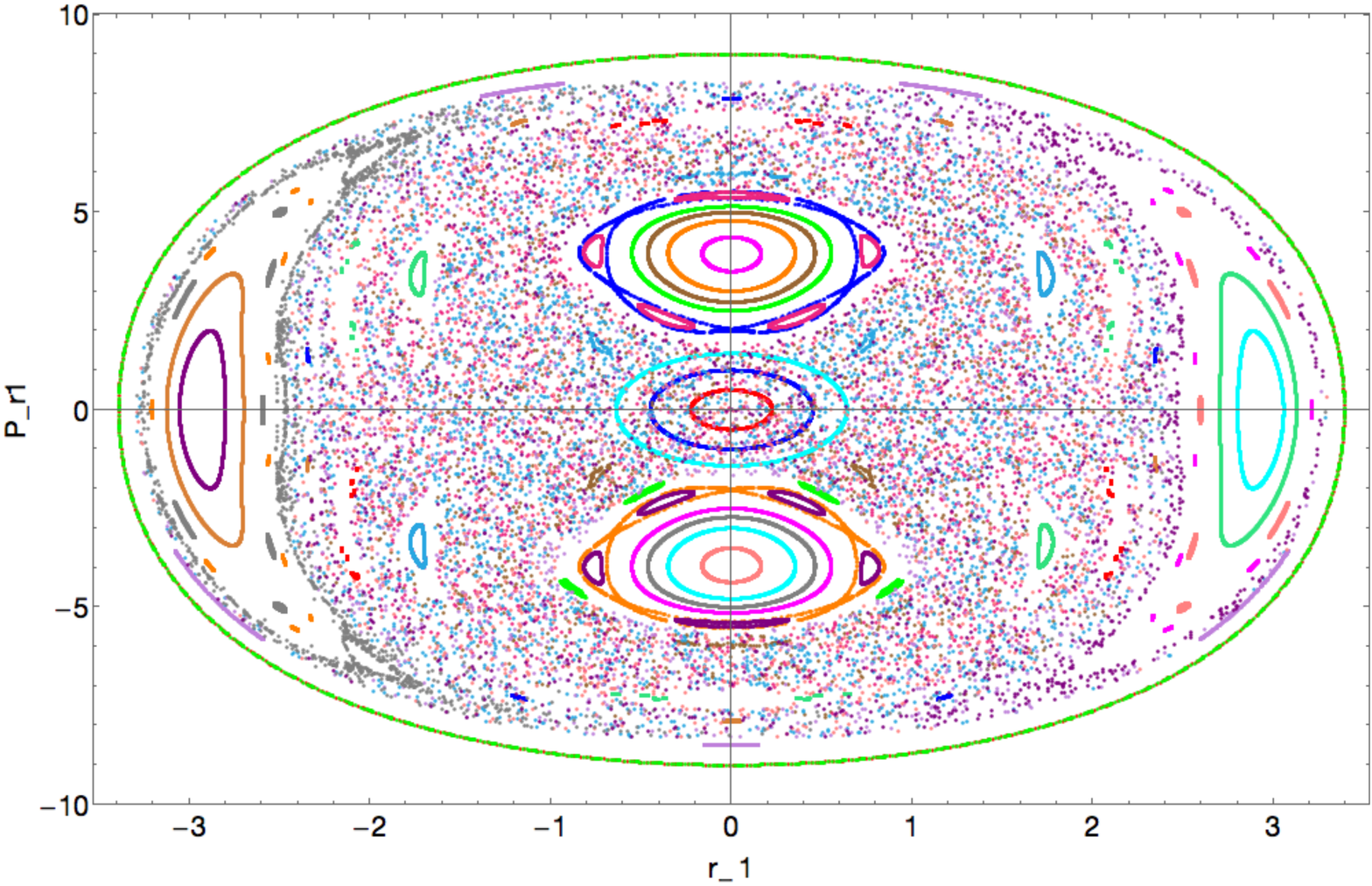} &
\includegraphics[scale=.25,angle=-0]{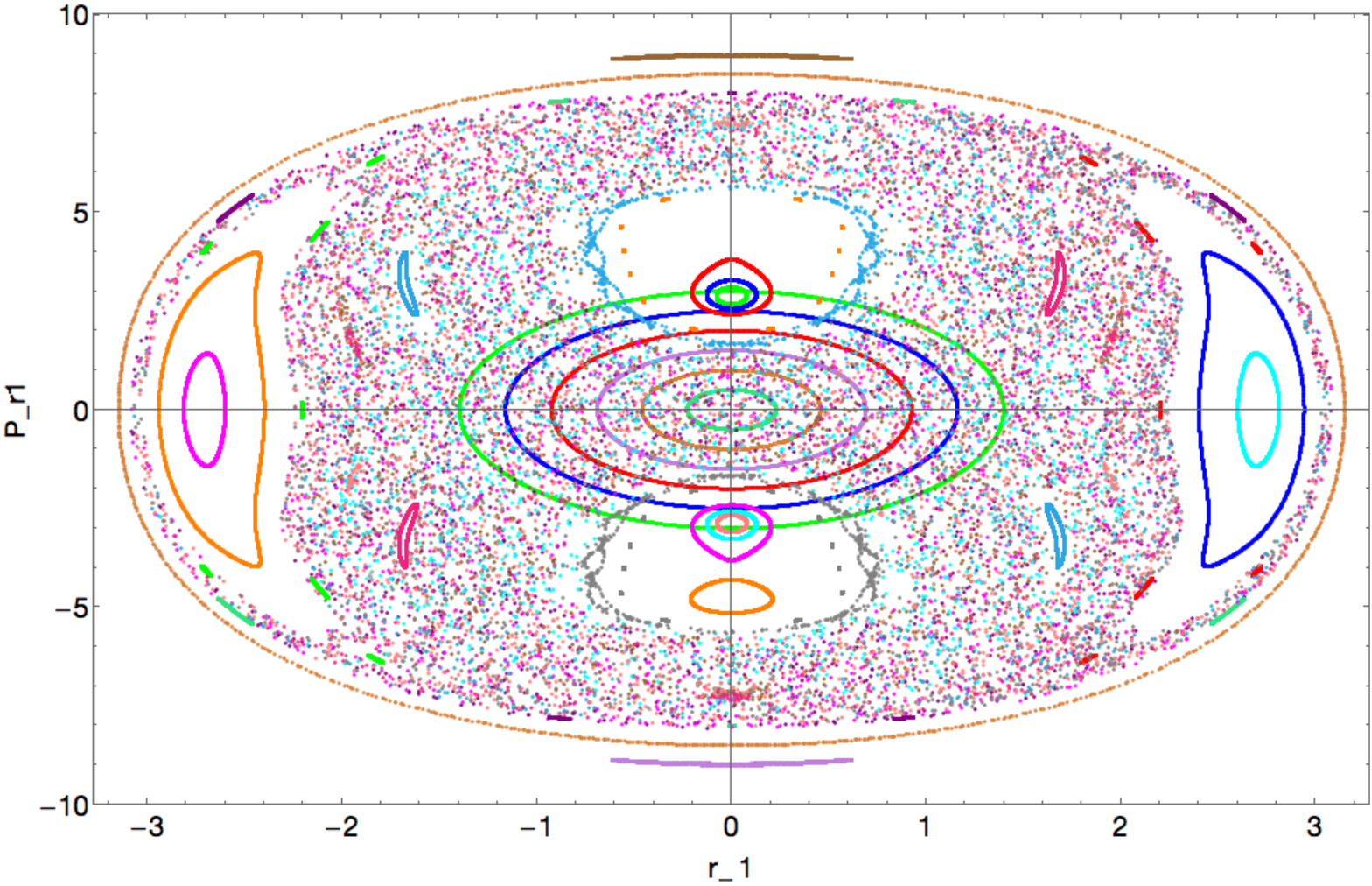} 
\vspace*{-0.1cm} 
\\
{\footnotesize (a) \quad Poincar\'e section with $E=1.0$} &
{\footnotesize (b) \quad Poincar\'e section with $E=5.0$} 
\vspace*{0.2cm}\\
\includegraphics[scale=.25,angle=-0]{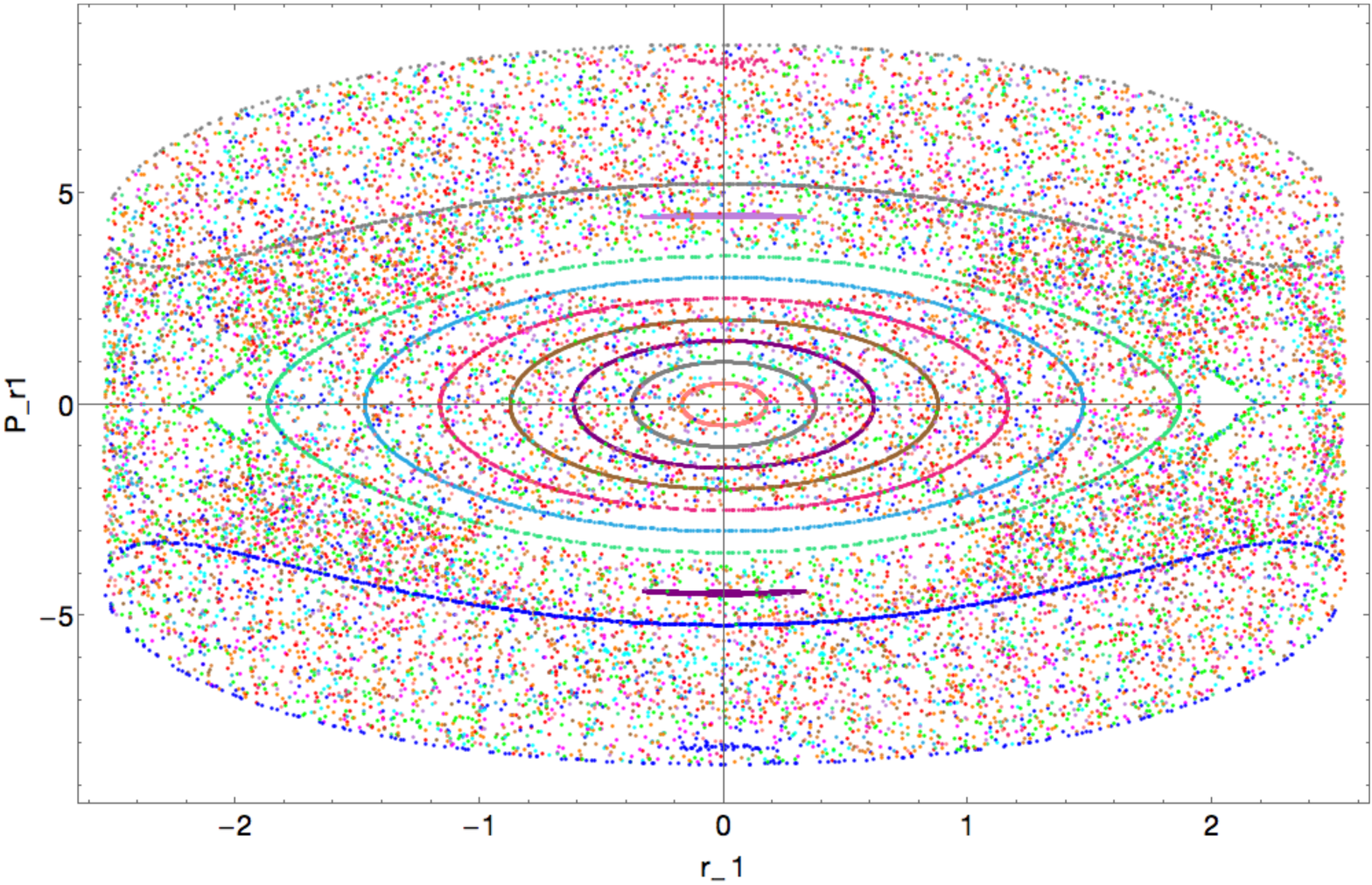} &
\includegraphics[scale=.2,angle=-0]{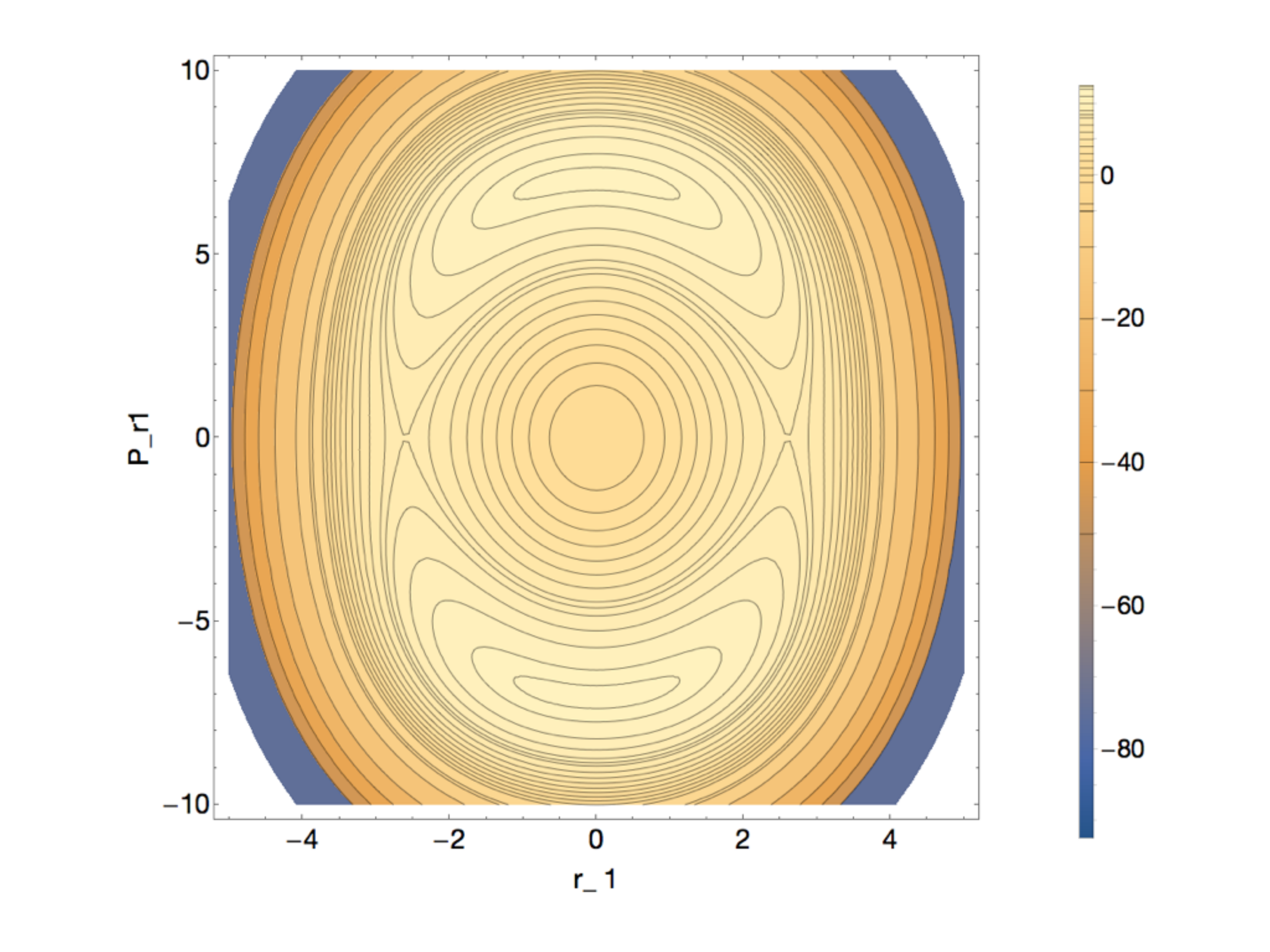}
\vspace*{-0.1cm} 
\\
{\footnotesize (c) \quad Poincar\'e section with $E=10$} &
{\footnotesize (d) \quad Energy contours in the $r_1$-$p_{r_1}$ phase space} \\
\end{tabular}
\caption{\label{Poincare-T11} \footnotesize Poincar\'e section 
with the ansatz (\ref{eq:ansatz_t11}). 
}
\end{center}
\end{figure}

Poincar\'e sections are plotted for $E=1.0$, 5.0 and 10 [Figs.\,\ref{Poincare-T11} (a)-(c)]. 
The sections are taken at $r_2 = 0$ with $p_{r_2} > 0$\,. 
The AdS radius $R$ and the winding numbers $\alpha_1$ and $\alpha_2$ 
are set to $R = 5.0$\,, $\alpha_1=2.0$ and  $\alpha_2=1.0$\,, respectively.  
The results clearly show that chaotic motions appear in each energy level,
and indicate that the near Penrose limit of AdS$_5\times T^{1,1}$ is also non-integrable.

\medskip 

It is worth mentioning the qualitative behavior of the Poincar\'e sections. 
The energy contours in the $r_1$-$p_{r_1}$ phase space at $r_2=p_{r_2}=0$ are drawn 
in Fig.\,\ref{Poincare-T11} (d)\,. 
A point is that it has a ring-like structure around the origin. 
In Fig.\,\ref{Poincare-T11} (a) chaotic motions have already appeared at $E=1.0$ 
together with islands and islets [Kolmogorov-Arnold-Moser (KAM) tori \cite{Ko,Ar,Mo}]. 
The location of islands is understood from the energy contours. 
When $E=5.0$\,, three islands collide each other and form a series of tori around the origin 
[Fig.\,\ref{Poincare-T11} (b)]. 
This position corresponds to the stable point around the origin in Fig.\,\ref{Poincare-T11} (d). 
When $E=10$\,, the centered tori grow up at last [Fig.\,\ref{Poincare-T11} (c)]. 

\medskip 

\begin{figure}[htbp]
\begin{center}
\includegraphics[scale=.25]{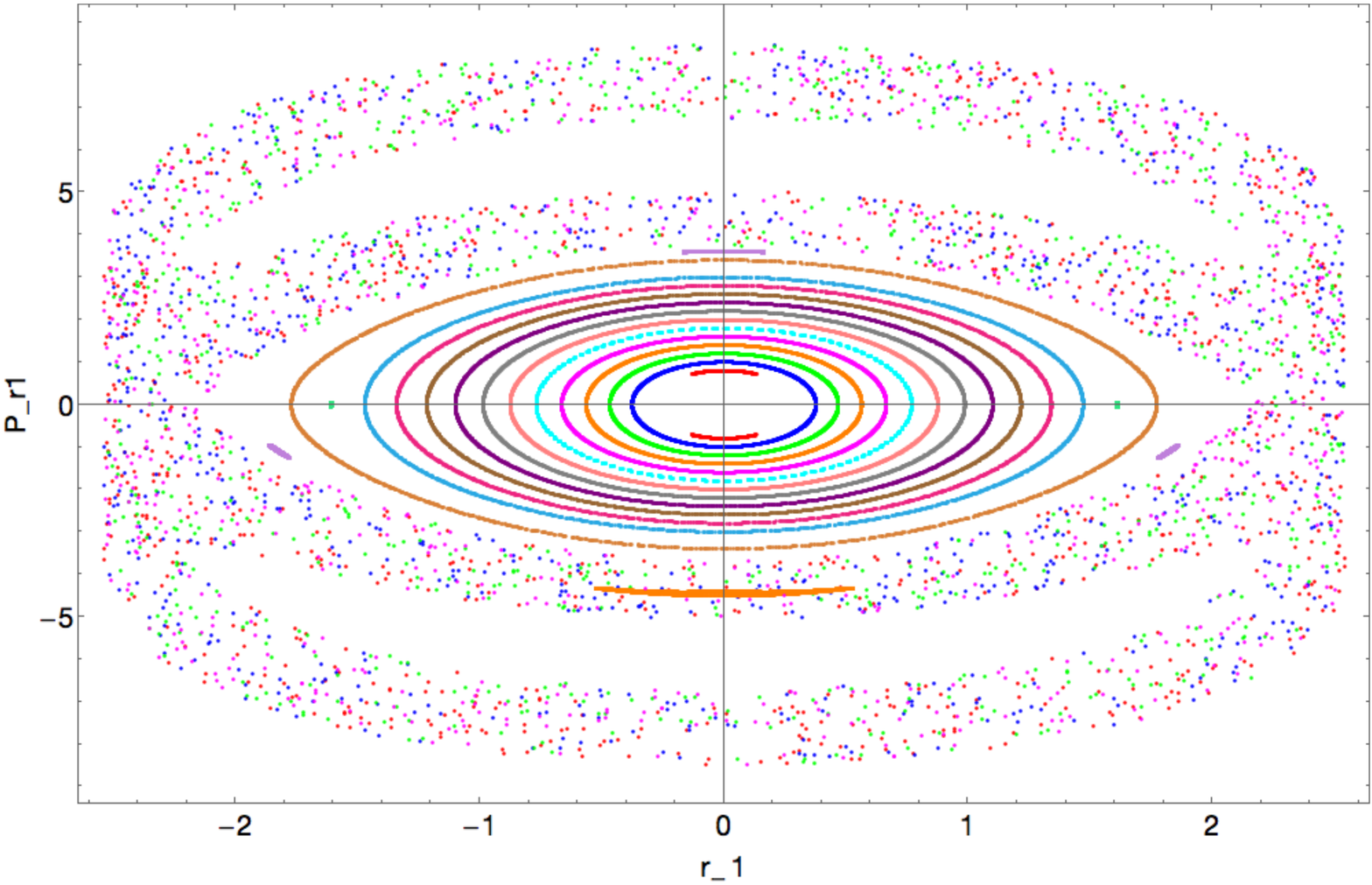} 
\vspace*{-0.2cm} \\
\caption{\label{another-Poincare} \footnotesize Improved Poincar\'e section with $E=10$\,. }
\end{center}
\end{figure}

Note that chaotic motions overlap with the tori in all of the sections. 
This is a peculiarity coming from the quartic terms of canonical momenta.  
Actually, we are not sure whether the Poincar\'e section we took here is suitable or not, 
though it should be enough to see the existence of chaos.  
There may be a possibility that another appropriate section can be chosen, for example, 
by imposing an additional condition to take the slice. 
For example, by taking a Poincar\'e section at $r_2=0$ with $0<p_{r_2}<5.2$\,, 
the overlap between chaotic trajectories and the KAM tori vanishes 
as plotted in Fig.\,\ref{another-Poincare}.

\medskip 

Finally, it is worth mentioning about Lyapunov spectra. 
The existence of the quartic terms of canonical momenta also makes it very difficult to compute them 
because the convergence of the exponents would become worse owing to it. 
In particular, this is the case even for the largest Lyapunov exponent.
So far, no satisfactory result has been obtained.

\subsection{No chaos in the radial direction of AdS$_5$}

In the previous subsection, we have observed chaotic motions associated with the chaos in $T^{1,1}$\,. 
As the next question, it may be interesting to ask the radial direction $r$ coming from the AdS$_5$ part 
may exhibit chaotic motions depending on initial conditions. In fact, the dynamics of $r$ is affected 
by the motions of the other $T^{1,1}$ variables and hence the answer would not be far from obvious. 

\medskip 

To answer this question, let us consider the following ansatz including the $r$-direction: 
\begin{eqnarray}
&  r = r(\tau)\,, \quad  p_r = p_r(\tau)\,,  \quad r_1 = r_1(\tau)\,, \quad \ p_{r_1} = p_{r_1}(\tau)\,, \quad 
r_2 = 0\,, \quad \ p_{r_2} = 0\,,  \no \\
& \quad \Phi_1 = \alpha_1\, \sigma\,, \quad p_{\Phi_1} = 0\,, \quad \Phi_2 = \alpha_2\, \sigma\,, 
\quad p_{\Phi_2} = 0\,. 
\label{eq:another_ansatz_t11}
\end{eqnarray}
Here $\alpha_i~(i=1,2)$ are winding numbers again. 

\medskip

The ansatz (\ref{eq:another_ansatz_t11}) simplifies the Hamiltonian (\ref{T1,1 Hamiltonian 0}) 
and (\ref{T1,1 Hamiltonian int}) as 
\begin{eqnarray}
  \mathcal{H}_0 &=& \frac{1}{2}\Bigl[ p_{r_1}^2 + p_{r_2}^2 
+r^2+ (1 + \alpha_1^2)\, r_1^2 \Bigr]\,, \no \\ 
 \mathcal{H}_{\rm int} &=& - \frac{1}{8} \Bigl[p_{r}^2 + p_{r_1}^2 -r^2 
+ (1 + \alpha_1^2)\,r_1^2 \Bigr]^2
+\frac{1}{6}r^4
+\frac{1}{2}\left(1-\alpha_1^2\right)\, r_1^4\,.
\label{hamiltonian_with_another_ansatz}
\end{eqnarray}
Note here that $\alpha_2$ does not appear.

\medskip

\begin{figure}[htbp]
\vspace*{0.2cm}
\begin{center}
\begin{tabular}{cc}
\includegraphics[scale=.25,angle=-0]{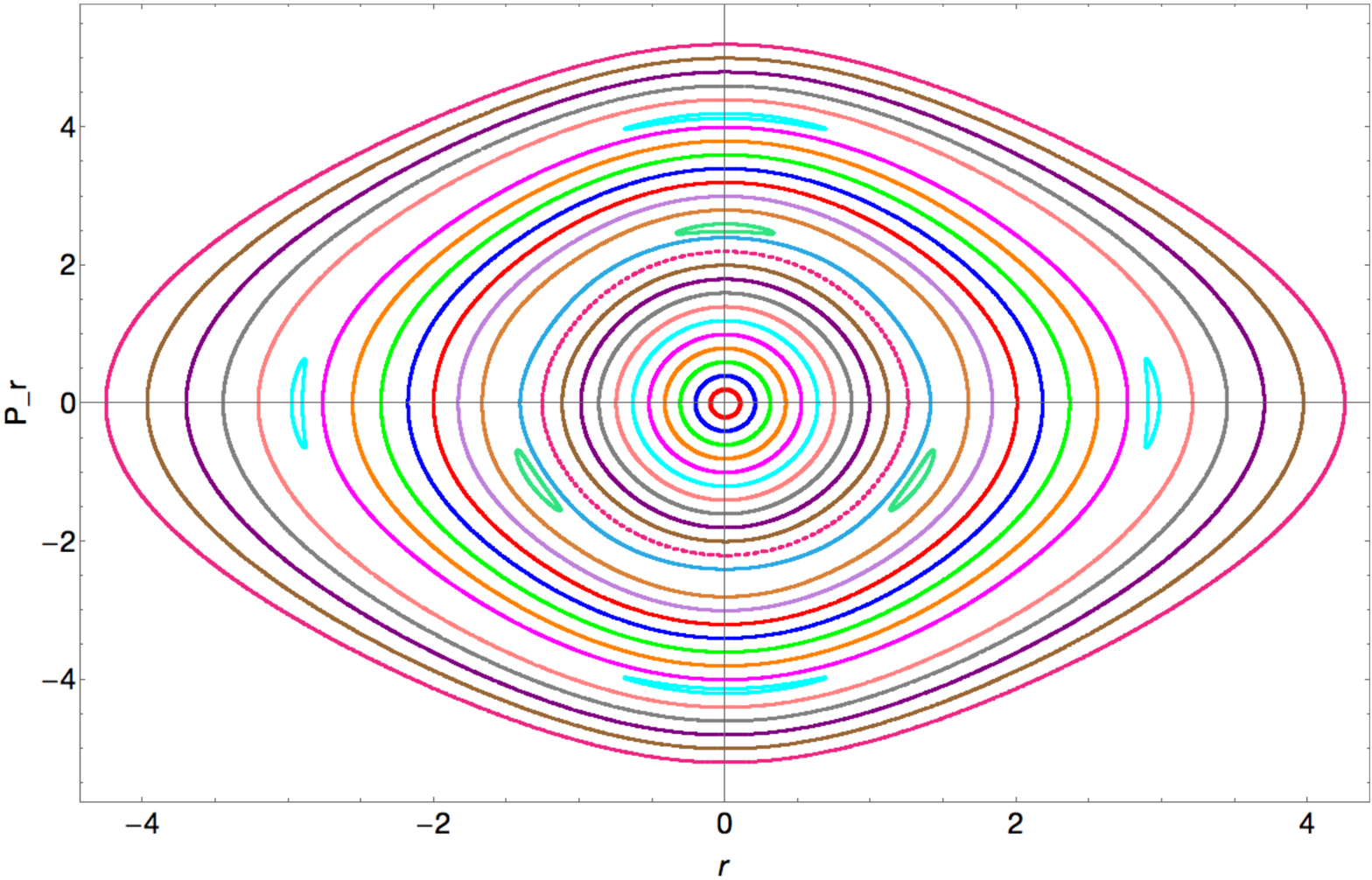} &
\includegraphics[scale=.2,angle=-0]{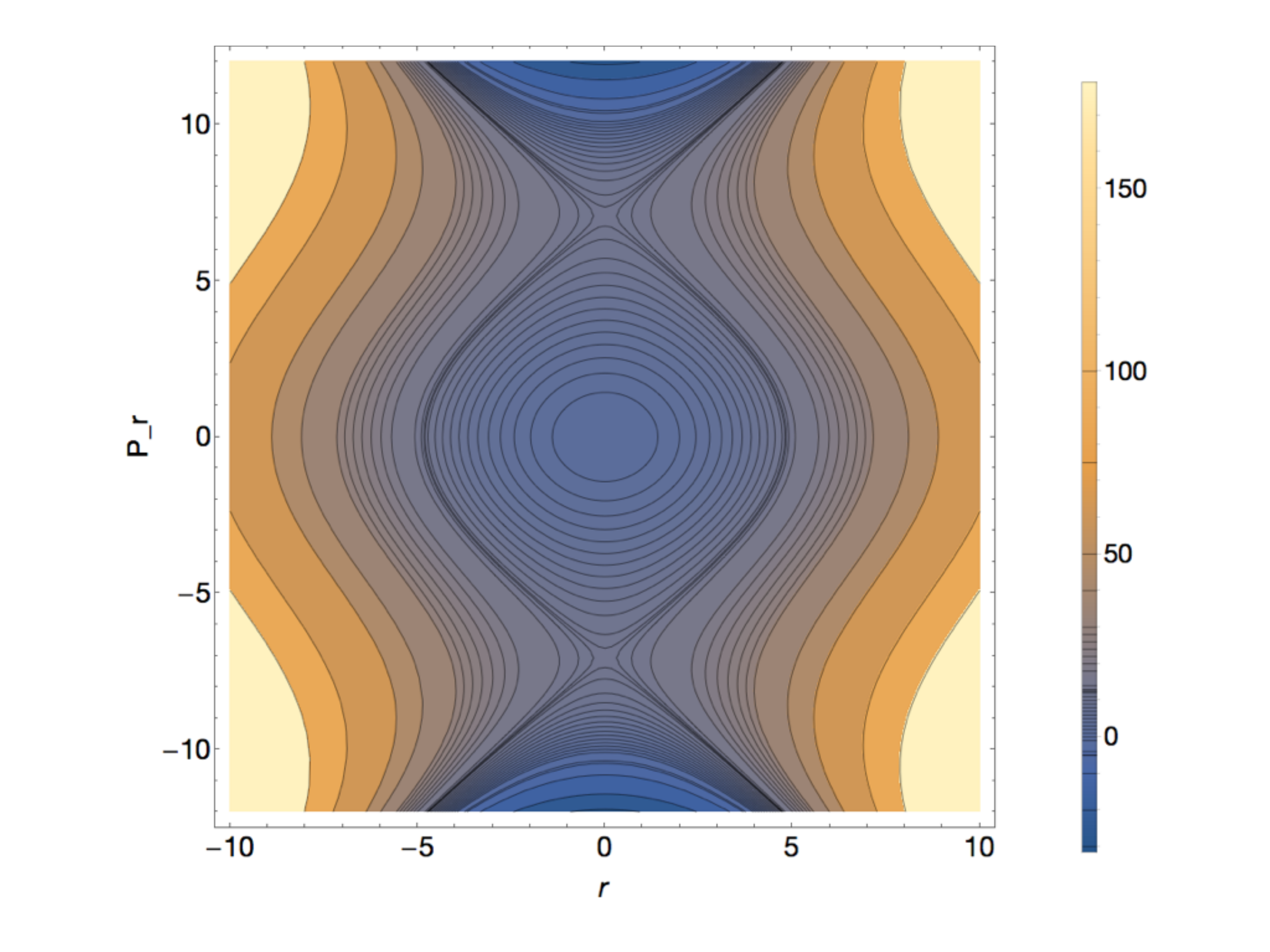} 
\vspace*{-0.2cm} 
\\
{\footnotesize (a) \quad Poincar\'e section with $E=10$} &
{\footnotesize (b) \quad Energy contours in the $r$-$p_{r}$ phase space} 
\\
\end{tabular}
\caption{\label{Poincare-another-T11} \footnotesize Poincar\'e section 
with the ansatz (\ref{eq:another_ansatz_t11}). 
}
\end{center}
\end{figure}

A Poincar\'e section at $r_1 = 0$ with $p_{r_1} > 0$ 
is plotted for $E=10$ with $R=\alpha_1=5.0$ [Fig.\,\ref{Poincare-another-T11} (a)]. 
Energy contours at $r_1=p_{r_1}=0$ are drawn in Fig.\,\ref{Poincare-another-T11} (b).
Figure \ref{Poincare-another-T11} (a) indicates that the KAM tori are not destroyed and 
there exist no chaotic motions for the $r$-direction. This is just an example, 
but we have obtained similar results for the other energy levels as far as we have tried. 
Thus, though we will not present a bunch of the plots, 
we have succeeded to give support for the classical integrability for the $r$-direction. 

\medskip 

The classical integrability should be associated with the integrability of AdS$_5$\,, but we have not obtained 
an analytical confirmation for this integrability. It may be a good direction to try to reveal it. 

\medskip

Finally, it should be remarked that the plot in Fig.\,\ref{Poincare-another-T11} (b) 
shows that the energy is not bounded for large values of $p_r$ 
and unbounded trajectories may appear. But the unbounded motions should not be confused 
with the onset of chaos\footnote{A simple exponential growth is not chaos. In general, the definition of chaos 
requires the finiteness of trajectories.}. 

\section{A near Penrose limit of AdS$_5\times$S$^5$ revisited} 

So far, we have considered the AdS$_5\times T^{1,1}$ case. 
Let us here revisit a near Penrose limit of AdS$_5\times$S$^5$\,.  
As a matter of course, the AdS$_5\times$S$^5$ geometry is known as an integrable background. 
However, it would be interesting to ask whether a near pp-wave limit of AdS$_5\times$S$^5$ 
is still integrable or not. This is because the interaction Hamiltonian contains the quartic terms of 
canonical momenta and it does not seem that the classical integrability is so obvious. 

\medskip 

First of all, let us introduce 
the metric of AdS$_5\times$S$^5$ with the global coordinates:  
\begin{eqnarray}
ds^2&=&R^2 (ds^2_{\rm AdS_5} + ds^2_{S^5})\,, \\ 
ds^2_{\rm{AdS_5}} &=&-\cosh^2 \rho\, dt^2 + d\rho^2 + \sinh^2 \rho\, d\Omega_3^2\,,  \\
ds^2_{S^5} &=&\cos^2\theta\,d\phi^2+d\theta^2+{\sin}^2\theta\,d{\Omega'_3}^2\,.
\end{eqnarray}
Here $R$ is the curvature radius of the AdS$_5$ and S$^5$\,.  
It is convenient to perform the coordinate transformations from $\rho$ and $\theta$ 
to $\tilde{z}$ and $\tilde{y}$ through the relations: 
\begin{eqnarray}
\cosh{\rho}=\frac{1+\tilde{z}^2/4}{1-\tilde{z}^2/4}\,, \qquad \cos{\theta}=\frac{1-\tilde{y}^2/4}{1+\tilde{y}^2/4}\,. 
\end{eqnarray}
Then the metric is rewritten as 
\begin{eqnarray}
ds^2_{\rm{AdS_5}} &=&-\left(\frac{1+\tilde{z}^2/4}{1-\tilde{z}^2/4}\right)^2dt^2
+\left(\frac{1-\tilde{y}^2/4}{1+\tilde{y}^2/4}\right)^2d\phi^2
+\frac{d\tilde{z}^2+ \tilde{z}^2 d\Omega_3^2}{(1-\tilde{z}^2/4)^2}
+\frac{d\tilde{y}^2+ \tilde{y}^2 d{\Omega'_3}^2}{(1+\tilde{y}^2/4)^2}\,.  \label{SO4}
\end{eqnarray}
In the metric (\ref{SO4})\,, the $SO(4)\times SO(4)$ isometry is manifest. 

\medskip 

Next, by following the work \cite{Schwarz}, the light-cone coordinates are introduced as 
\begin{eqnarray}
\tilde{x}^+ &=& t\,, \qquad \tilde{x}^- = -t+\phi\,.
\end{eqnarray}
After rescaling the coordinates as 
\begin{equation}
\tilde{x}^+=x^+\,, \quad \tilde{x}^-=\frac{{x}^-}{R^2}\,, \quad \tilde{z}=\frac{z}{R}\,, 
\quad \tilde{y}= \frac{y}{R}\,,
\end{equation}
the $R \to \infty$ limit is taken. 
The resulting metric is given by
\begin{eqnarray}
ds^2&=&{ds_0}^2+\frac{1}{R^2}{ds_2}^2+{\cal O}\left( \frac{1}{R^4}\right)\,, \label{near} \\
{ds_0}^2&=&2dx^+dx^--(z^2+y^2)(dx^+)^2+dz^2+z^2d{\Omega_3}^2+dy^2+y^2d{\Omega'_3}^2\,, \\
{ds_2}^2&=&-2y^2dx^+dx^-+\frac{1}{2}(y^4-z^4)(dx^+)^2+(dx^-)^2\no \\
&& \quad +\frac{1}{2}z^2\left(dz^2+z^2d{\Omega_3}^2\right)-\frac{1}{2}y^2\left(dy^2+y^2d{\Omega'_3}^2\right)\,.
\end{eqnarray}
This metric with the sub-leading corrections was originally discussed in \cite{Schwarz}. 

\medskip 

Now it is an easy task to derive the light-cone Hamiltonian ${\cal{H}}_{\rm lc}$ 
on the background (\ref{near}) by making use of (\ref{light cone Hamiltonian}).
After setting $p_-=1$ and dropping a constant term, we obtain the Hamiltonian:
\begin{eqnarray}
{\cal{H}}_{\rm lc}&=&{\cal{H}}_0+\frac{1}{R^2}{\cal{H}}_{\rm int}+{\cal O}\left( \frac{1}{R^4}\right)\,, 
\label{Hlc-S5} \\
{\cal{H}}_0&=&\frac{1}{2}\left((p_A)^2+({x'}^A)^2+(x^A)^2\right)\,, \\
{\cal{H}}_{\rm int}&=&\frac{1}{4}\left(z^2({p_y}^2+{y'}^2+2{z'}^2)-y^2({p_z}^2+{z'}^2+2{y'}^2)\right)\no \\
&& \qquad +\frac{1}{8}\left(((x^A)^2)^2-((p_A)^2+({x'}^A)^2)^2\right)+\frac{1}{2}(p_A{x'}^A)^2\,, 
\end{eqnarray}
where $x^A=(z,y)$ and $p_A=(p_z,p_y)$\,.
Here we have assumed that a constant position is taken in each of two S$^3$'s,
and dropped the terms concerned with ${d\Omega_3}^2$ and ${d\Omega'_3}^2$\,, as we did in Sec.\ 3.
In addition, we will not consider the higher-order terms with $\mathcal{O}(1/R^4)$\,.

\subsubsection*{Poincar\'e section}

The next task is to investigate numerically the dynamics of the Hamiltonian system with (\ref{Hlc-S5})\,. 
In the following, we will compute a Poincar\'e section 
and provide support for the classical integrability of the system with (\ref{Hlc-S5})\,. 

\medskip 

To make the system simpler, let us take the following ansatz, 
\begin{eqnarray}
  \label{eq:ansatz_s5}
  & y = y(\tau)\,, \quad \ p_{y} = p_{y}(\tau)\,, \quad 
z = z(\tau)\,, \quad \ p_{z} = p_{z}(\tau)\,.
\end{eqnarray}
With this ansatz, the light-cone Hamiltonian is simplified as 
\begin{eqnarray}
{\cal{H}}_{\rm lc}&=&{\cal{H}}_0+\frac{1}{R^2}{\cal{H}}_{\rm int}+{\cal O}\left( \frac{1}{R^4}\right)\,, \label{Hlc-S5-ansatz} \\
{\cal{H}}_0&=&\frac{1}{2}\left((p_A)^2+(x^A)^2\right)\,, \\
{\cal{H}}_{\rm int}&=&\frac{1}{4}\left(z^2{p_y}^2-y^2{p_z}^2\right)+\frac{1}{8}
\left(\bigl((x^A)^2\bigr)^2-\bigl((p_A)^2\bigr)^2\right)\,.
\label{S^5 Hamiltonian}
\end{eqnarray}

\begin{figure}[htbp]
\begin{center}
\begin{tabular}{cc}
\includegraphics[scale=.25]{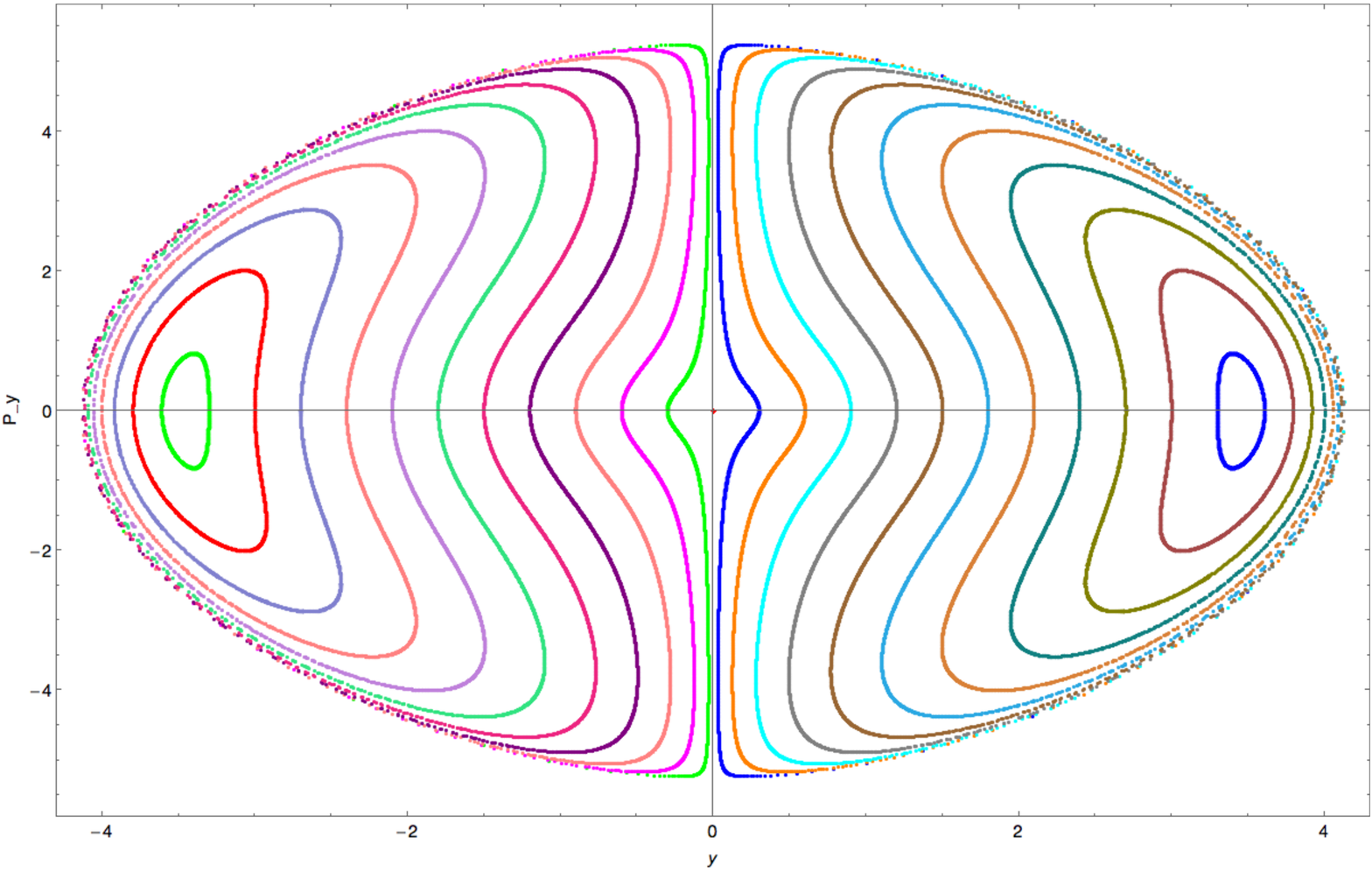} &
\includegraphics[scale=.2]{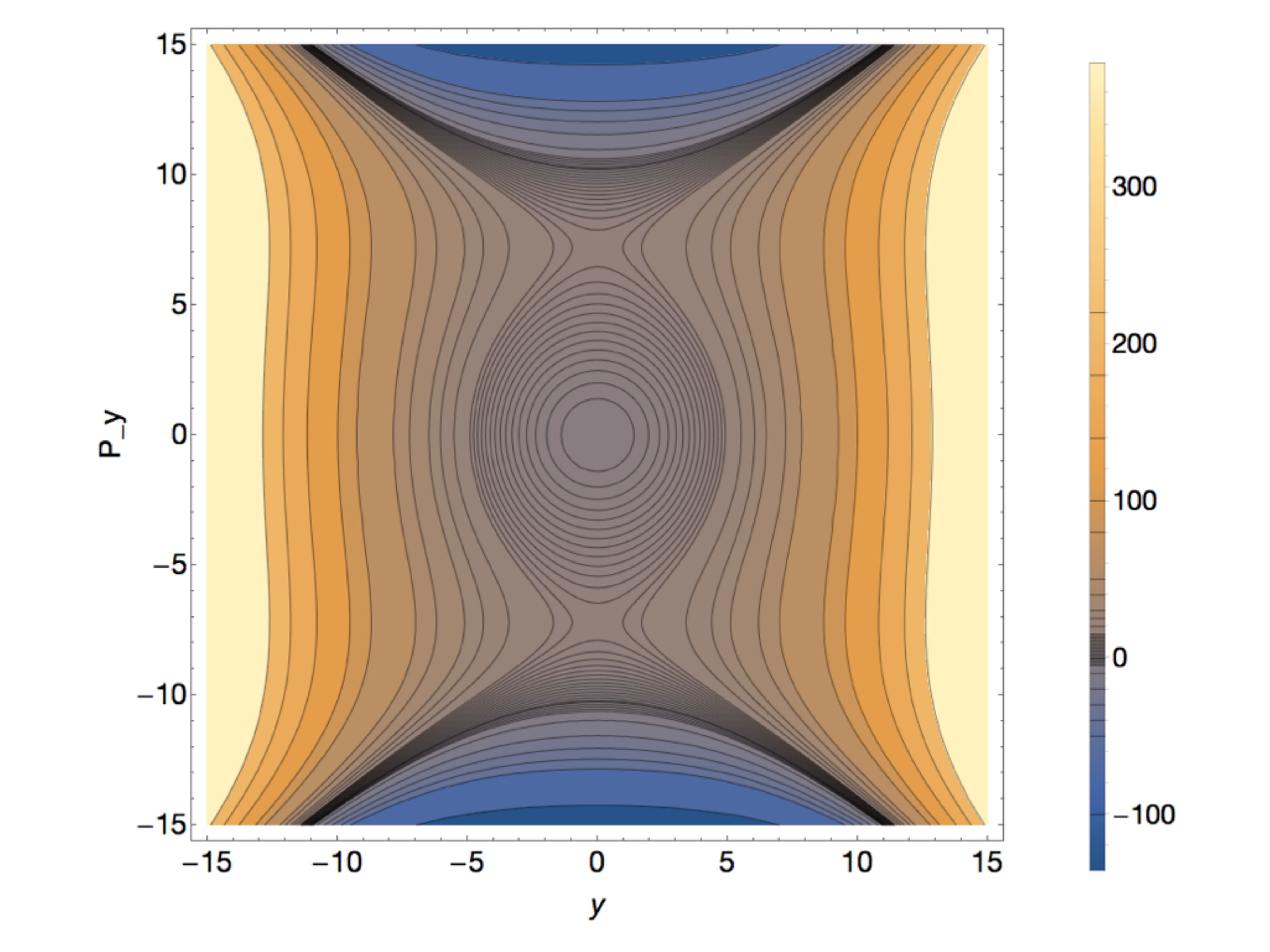} 
\vspace*{-0.2cm} \\
{\footnotesize (a) \quad Poincar\'e section with $E=10$} &
{\footnotesize (b) \quad Energy contours in the $y$-$p_y$ phase space}  \\
\end{tabular}
\caption{\label{Poincare-S_5} \footnotesize Poincar\'e section with the ansatz (\ref{eq:ansatz_s5}). }
\end{center}
\end{figure}

A Poincar\'e section is presented in Fig.\,\ref{Poincare-S_5} (a)\,,  
The section is taken at $z = 0$ with $p_{z} > 0$ and 
computed for $E =10$ with $R = 5.0$\,. Energy contours are drawn in the $y$-$p_{y}$ phase space 
with $z=p_{z}=0$ [Fig.\,\ref{Poincare-S_5} (b)].
Figure \ref{Poincare-S_5} (a) shows that there are no chaotic motions at $E=10$\,. 
Although the energy is sufficiently high, the KAM tori are not destroyed. 
Note that the plot in Fig.\,\ref{Poincare-S_5} (b) shows that the energy is not bounded for large values of $p_y$ again. 
But, as in Sec.\ 2.2, the unbounded motions should not be interpreted as the onset of chaos. 

\medskip 
  
Figure \ref{Poincare-S_5} (a) is just an example, but beautiful KAM tori continue to survive for other energy levels, 
as far as we have tried. Thus, though we will not present a bunch of the plots,  
we have obtained support for the classical integrability even in the near Penrose limit of 
AdS$_5\times$S$^5$\,. 

\medskip 

This result should be related to the classical integrability of type IIB string theory on AdS$_5\times$S$^5$ \cite{BPR}. 
Then, at least in principle, it would be possible to show the integrability by explicitly 
constructing an infinite number of conserved charges or the Lax pair. 
However, because of the quartic terms of canonical momenta, it seems quite difficult 
and hence our numerical support would be valuable.

\section{Conclusion and Discussion}

In this paper, we have considered chaotic motions of a classical string in a near Penrose limit of AdS$_5\times T^{1,1}$\,. 
We have shown that sub-leading corrections in a Penrose limit provide an unstable separatrix, 
so that chaotic motions are generated as a consequence of collapsed KAM tori. 
By deriving a reduced system composed of two degrees of freedom with a winding string ansatz, 
we have computed Poincar\'e sections and provided support for the existence of chaos. 
In addition, we have argued that no chaos appears in a near Penrose limit of AdS$_5\times$S$^5$\,, as expected 
from the classical integrability of the parent system. 

\medskip 

There are some open problems associated with the chaos in the AdS$_5\times T^{1,1}$\,.  
A most important issue is to clarify what kind of gauge-theory operators 
correspond to the chaotic string solutions. We have studied here a near Penrose limit 
and hence the associated operators should be almost BPS. That is, a few impurities are inserted into 
the BPS vacuum operator. Hence it seems likely that 
the problem would now be much easier than the setup discussed in \cite{T11}, 
because the associated composite 
operators are quite intricate in the case of the full geometry.
However, we have no definite answer for the operators so far. We need to make more of an effort,  
For example, by following the argument for a near Penrose limit of AdS$_5\times$S$^5$ \cite{Schwarz}.
It may also be useful to try to figure out a fractal structure associated with the chaos. 
Probably, one may expect that the impurities would randomly be inserted in the vacuum operator.

\medskip 

We hope that our result would open up a new arena to check the AdS/CFT correspondence 
even for chaotic string solutions.

\subsection*{Acknowledgments}

It is a pleasure to acknowledge helpful discussions with Jun-ichi Sakamoto  
and Shin-ichi Sasa. 
The work of D.~K.\ is supported by the Japan Society for the Promotion of Science (JSPS).
The work of K.~Y.\ is supported by Supporting Program for  Interaction-based Initiative Team Studies 
(SPIRITS) from Kyoto University and by the JSPS Grant-in-Aid for Scientific Research (C) No.~15K05051. 
This work is also supported in part by the JSPS Japan-Hungary Research Cooperative Program 
and the JSPS Japan-Russia Research Cooperative Program.

\end{document}